\documentstyle{elsart}
\begin{document}

 \newcommand{\sectlabel}[1]{}
 \newcommand{\eqlabel}[1]{\label{eq.#1}}
 \newcommand{\eq}[1]{(\ref{eq.#1})}	
 \renewcommand\medskip{}

\begin{frontmatter}

 \title{Why quantum bit commitment and \\ideal quantum coin tossing are
impossible.
}
 \author{Hoi-Kwong Lo\thanksref{HP}} and \author{H. F. Chau\thanksref{HKU}}
 \address{School of Natural Sciences, Institute for Advanced Study, Princeton,
  NJ 08540}
\thanks[HP]{Address from July 1, 1997:
Hewlett-Packard Labs, Filton Road, Stoke Gifford, Bristol BS12 6QZ, UK.~
email: hkl@hplb.hpl.hp.com}
\thanks[HKU]{Address from July 1, 1996:
Department of Physics, University of Hong Kong, Pokfulam Road,
Hong Kong.~~email: hfchau@hkusua.hku.hk}
 \date{10 May, 1996\\
Revised, 21 Nov., 1997}

 \begin{abstract}
 There had been well known claims of unconditionally secure quantum
protocols for bit commitment.  However, we, and
independently Mayers, showed that all {\it proposed} quantum bit commitment
schemes are, in principle, insecure because the
sender, Alice, can almost always cheat successfully by using an
Einstein-Podolsky-Rosen (EPR) type of attack
and delaying her measurements.  One might wonder if secure quantum bit
commitment protocols exist at all.  We answer this question by
showing that
the same type of attack by Alice will, in principle, break {\em any}
bit commitment scheme. The cheating strategy generally requires a quantum computer.
We emphasize the generality of this ``no-go theorem'':
Unconditionally secure bit commitment schemes based
on quantum mechanics---fully quantum, classical or quantum but with
measurements---are all ruled out by this result.
Since bit commitment is a useful primitive for building up
more sophisticated protocols such as zero-knowledge proofs, our results cast
very serious doubt on the security of quantum cryptography in the so-called
``post-cold-war'' applications.
We also show that {\it ideal} quantum coin tossing is impossible because of
the EPR attack. This no-go theorem for ideal quantum coin tossing
may help to shed some lights on the possibility of non-ideal protocols.

 \end{abstract}

 \end{frontmatter}

\section{Introduction}\sectlabel{intro}

Quantum cryptography was first proposed by Wiesner \cite{Wiesner} more than
two decades ago in a paper that remained unpublished until 1983. Recently,
there have been lots of renewed activities in the subject. The most
well-known application of quantum cryptography is key distribution
\cite{BB84,Bennett:SciAm,Ekert}.  The aim of key distribution is to allow two
users to generate a shared random string of information that can, for
example, be used to make their messages in subsequent communication totally
unintelligible to an eavesdropper.  Quantum key distribution is generally
believed to be secure
\cite{Bennett:Exp,Mor,Deutsch,Mayers1,Mayers3} because, it is impossible (for
an eavesdropper) to make copies (or clones) of non-orthogonal states in
quantum mechanics without violating unitarity. Moreover, measuring a quantum
system generally disturbs it because quantum mechanical observables can be
non-commuting. For this reason, eavesdropping on a quantum communication
channel will generally leave unavoidable disturbance in the transmitted
signal which can be detected by the legitimate users.

\medskip

This paper does not concern quantum key distribution. However, it concerns a
class of more fancy applications of quantum cryptography that have also been
proposed \cite{Ar:bit,Ar:ot,BB84,BBCS92,BC91,BCJL93}.  Those applications are
probably more useful in the post-cold-war era.  A typical problem in
``post-cold-war'' quantum cryptography is the two-party secure computation,
in which both parties would like to know the result of a computation but
neither side wishes to reveal its own data. For example, two firms will
embark on a joint venture if and only if their combined capital available for
the project is larger than one million dollars. They would like to know if
this condition is fulfilled but neither wishes to reveal the exact amount of
capital it commits to the project.  In classical cryptography, this can be
done either through trusted intermediaries or by invoking some unproven
computational assumptions such as the hardness of factoring. The big question
is whether quantum cryptography can get rid of those requirements and achieve
the same goal using the laws of physics alone.

\medskip

Until recently, there had been much optimism in the subject.  Various
protocols for say bit commitment, coin tossing and oblivious transfer of
quantum cryptography had been proposed
\cite{Ar:bit,Ar:ot,BB84,BBCS92,BC91,BCJL93}.  In particular, the BCJL
\cite{BCJL93} bit commitment scheme had been claimed to be provably
unbreakable.  It should be noted that bit commitment is a crucial primitive
in building up more sophisticated protocols: It has been shown by Yao
\cite{Yao} that a secure quantum bit commitment scheme can be used to
implement a secure quantum oblivious transfer scheme whereas Kilian
\cite{Kilian} has shown that, in classical cryptography, oblivious transfer
can be used to implement secure two-party computations.  This chain of
arguments, therefore, seems to suggest that quantum bit commitment alone is
sufficient for implementing secure two-party computations, thus solving a
long standing problem in cryptography.  However, this widespread belief in
the power of post-cold-war quantum cryptography has recently been seriously
challenged by the independent works of ours \cite{Lo2} and of Mayers
\cite{Mayers2} which showed that, contrary to popular belief, all {\it
proposed} quantum bit commitment schemes are, in fact, insecure.
(The non-ideal case was demonstrated rigorously by Mayers.)
This is
because the sender, Alice, can always cheat successfully by using an EPR-type
of attack and delaying her measurement until she opens her commitment.

\medskip

The result of Mayers and ours, however, left open the possibility of the
existence of some yet unknown secure quantum bit commitment scheme.  This
paper eliminates this possibility: We show that, provided that a cheater has
a quantum computer, unconditionally secure quantum bit commitment is
impossible. This is because an EPR-type
of attack by at least one of the users can almost always break all bit commitment
schemes. We
emphasize the generality of this
no-go theorem. It applies to {\it any} scheme that is consistent
with the laws of quantum mechanics. It does not matter whether the scheme is
purely quantum, classical or quantum but with measurements.

The reasoning behind its generality is that any two-party quantum computation
(possibly with measurements at the intermediate steps) can be
{\it rephrased} into a fully
quantum scheme where measurements are delayed until the very end of the computation.
This is chiefly because, by definition, quantum mechanics should
provide the most fundamental description of the evolution part of such
a computation and, therefore, measurements can be
taken into account only at the very end.

More precisely, we explicitly construct a cheating strategy that
will allow Alice to cheat successfully against Bob {\it even if\/} Bob
has a quantum computer. Now any procedure (including measurements,
attachment of ancilla, unitary transformations) followed by Bob can be
rephrased into one in which Bob does have a quantum computer but just
fails to make full use of it. Therefore, this `sure-win' strategy will
allow Alice to defeat any Bob. Consequently, unconditionally secure
bit commitment based on
quantum mechanics is impossible.

We also demonstrate the
insecurity of {\it ideal} quantum coin tossing against the same type of attack.

\medskip

We acknowledge the receipt of a preprint of Dominic Mayers about the
impossibility of quantum bit commitment. This preprint contains the essential
result and approach to bit commitment that we present here except that, in
our opinion, it does not define in sufficient detail the general model that
it uses for quantum protocols and therefore the model is too vague.  To
answer the question in a more satisfactory manner and to make the discussion
more concrete and comprehensive, we strongly feel the need to use a variant
of the Yao's model.  Besides, such a concrete model is essential for our
discussion of the impossibility of ideal quantum coin tossing.

\section{Quantum bit commitment}\sectlabel{bit}

A general bit commitment scheme involves two parties, a sender Alice and a
receiver, Bob. Suppose that Alice has a bit ($b = 0$ or $1$) in mind, to
which she would like to be committed towards Bob. That is to say, she wishes
to provide Bob with a piece of evidence that she has a bit in mind and that
she cannot change it. Meanwhile, Bob should not be able to tell from that
evidence what $b$ is. At a later time, however, it must be possible for Alice
to {\it open} the commitment. That is, Alice must be able to show Bob which
bit she has committed to and convinced him that this is indeed the genuine
bit that she had in mind when she committed.\footnote{A bit commitment scheme
is useful for say implementing a coin tossing scheme. See footnote~9 below.}

\medskip

In a simple implementation of bit commitment, Alice writes down her bit in a
piece of paper, puts it in a box and locks the box.  She then sends the
locked box to Bob, but keeps the key.  The whole point is that while Bob
cannot open the box to learn the value of the bit without the key, Alice
cannot change her mind. At a later time, Alice gives the key to Bob, who
opens it and verifies the value of the bit. This implementation relies solely
on the physical security of the box and is, thus, infeasible in the
electronic age.

\medskip

What constitutes to a cheating by Alice? If Alice commits to a particular
value of $b$ (e.g., $b=0$) during the commitment phase and attempts to change
it to another value (e.g., $b=1$) during the opening phase, Alice is
cheating. A bit commitment scheme is secure against Alice only if such a fake
commitment will be discovered by Bob.  In this section, we show that,
contrary to popular belief, all quantum bit commitment schemes are, in
principle, insecure against a cheating Alice.

\subsection{Model of two-party quantum protocols}\sectlabel{two}

Quantum bit commitment and coin tossing are examples of two-party quantum
protocols.  A two-party quantum protocol involves a pair of quantum machines
in the hands of two users, $A$ (Alice) and $B$ (Bob) respectively, which
interact with each other through a quantum channel, $C$.  More formally, we
consider the direct product $H$ of the three Hilbert spaces $H_A$, $H_B$ and
$H_C$ where $H_A$ ($H_B$) is the Hilbert space of Alice's (Bob's) machine and
$H_C$ is the Hilbert space of the channel. We assume that initially each
machine is in some specified pure quantum state. $A$ and $B$ then engage in a
number of rounds of quantum communications with each other through the
channel $C$. More concretely, each of $A$ and $B$ alternately performs a unitary
transformation on $H_D \otimes H_C$ where $D \in \{A,B\}$.
 
\medskip

The above model is a simplification of a model proposed by Yao
\cite{Yao}. Although Yao apparently did not emphasize the generality of his
model, it appears to us that any realistic two-party computation can be
described by Yao's model. For instance, since Alice and Bob are separated by
a long distance, it is impractical to demand simultaneous two-way
communications between them.\footnote{More importantly, simultaneity has no
invariant meaning in special relativity.}  The idea of alternate rounds of
one-way communications in Yao's model is, therefore, reasonable.  However,
there are two significant distinctions between Yao's model and ours. First,
Yao's model deals with mixed initial states whereas we assume that the
initial state of each machine is pure. Second, in Yao's model, the user $D$
does two things in each round of the communication: $D$ carries out a
measurement on the current mixed state of the portion of the space, $H_D
\otimes H_C$, in his/her control and then performs a unitary transformation
on $H_D \otimes H_C$.  In our model, the measurement step has been
eliminated.

\medskip

Here we argue that our simplifications are desirable for our current
analysis. Let us consider the first distinction.  In assuming that the
initial state of each machine is pure, we are just giving the users complete
control over the initial states of the machines. Any situation with mixed
initial state can be included in our consideration simply by attaching a
quantum dice to a machine and considering the pure state as describing the
combined state of the two.

\medskip

What about the second distinction?  Essentially, we give Alice and Bob
quantum computers and quantum storage devices.  Therefore, they can execute a
quantum bit commitment scheme by unitary transformations.

\subsection{The key observation}

The reader may still wonder about the role of measurement. Does it mean that
one should follow all possible histories of the outcomes and consider {\it
decoherence}? Is our proof really general?  We argue that our proof is
general. The key insight is the following: To show that {\it all\/} bit
commitment schemes (classical, quantum or quantum but with some measurements)
are insecure, it suffices to consider {\it only} a fully quantum bit
commitment scheme where both Alice and Bob have quantum computers. This is
because {\it any} other procedure followed by Bob in a bit commitment scheme
can be rephrased as a quantum bit commitment scheme where Bob does have a
quantum computer but just fails to make full use of it.

\medskip

Now, since we will show that Alice has a winning strategy against Bob even if
he makes full use of his quantum computer, it is clear that this ``sure-win''
strategy by Alice will defeat a Bob who fails to make full use of his quantum
computer. Therefore, the insecurity of a fully quantum bit commitment scheme
automatically implies the insecurity of all bit commitment schemes (purely
quantum, classical or quantum scheme but with measurements).

\medskip

The above remark --- that it is unnecessary to consider decoherence and
measurements --- applies also to our discussion of ideal coin tossing. Such a
unitary description will greatly simplify our discussion.\footnote{We thank
L. Goldenberg and D. Mayers for a discussion on the generality of Yao's
model.}

\subsection{Procedure of quantum bit commitment}

The most general procedure for a quantum bit commitment scheme can be
rephrased in the following unitary description.

\medskip

(a) Preparation of states: Alice chooses the value of a bit $b$ to which she
would like to be committed towards Bob. If $b=0$ (respectively $b=1$), she
prepares a state $ |0 \rangle$ (respectively $ |1 \rangle$) for $H_A$.  Bob
prepares a state $|BC \rangle$ for the product Hilbert space $H_B \otimes
H_C$.  All the states $ |0 \rangle$, $ |1 \rangle$ and $|BC \rangle$ are
specified by the protocol and are known to both Alice and Bob.

\medskip

(b) The actual commitment: Step (b) involves a specified and fixed number of
rounds of quantum communication alternately between Alice and Bob.  As noted
above, each round of quantum communication can be modeled as a unitary
transformation on $H_D \otimes H_C$ ($D \in \{A,B\}$), which in turn induces
a unitary transformation on the space $H= H_A \otimes H_B \otimes H_C$.

\medskip

Notice that for an {\em ideal} bit commitment, it must be the case that, at
the end of step (b), Bob still has absolutely no information about the value
of the committed bit $b$. (We will relax this assumption when we come to the
non-ideal case in the next subsection.)  Now that the commitment has been
made, both sides may wait an arbitrary length of time until the last step:

\medskip

(c) Opening of the commitment: A specified and fixed number of rounds of
quantum communication alternately between Alice and Bob are again involved.
As in step (b), we model each round of quantum communication as a unitary
transformation on $H_D \otimes H_C$ ($D \in \{A,B\}$), which in turn induces
a unitary transformation on the space $H= H_A \otimes H_B \otimes H_C$.

\medskip

In a secure bit commitment scheme, Bob will learn the value of $b$ and be
convinced that Alice has already committed to that value of $b$ at the end of
step (b) and cannot change it anymore in step (c).

\medskip

In what follows, we argue that the above general scheme necessarily fails
because Alice can always cheat successfully by using {\em reversible} unitary
operations in step (b) and subsequently rotating a state that corresponds to
$b=0$ to one that corresponds to $b=1$ and vice versa in the beginning of
step (c).  Note that, for Alice to cheat, she generally needs
a quantum computer to perform the desirable unitary transformation.

\medskip

Let us justify our claim. Consider more closely the situation at the end of
step (b), the commitment phase.  Let $|0 \rangle_{\rm com} $ and $|1
\rangle_{\rm com} $ denote the state of $H = H_A \otimes H_B \otimes H_C$ at
that time corresponding to the two possible values of $b$ respectively. In
order that Alice and Bob can follow the procedures, they must know the exact
forms of all the unitary transformations involved.\footnote{As stated
earlier, any probabilistic scheme can be rephrased as a deterministic one by
considering the state of the combined system of the quantum dice and the
original system.}  Therefore, Alice must be capable of computing the two
states $|0 \rangle_{\rm com} $ and $|1 \rangle_{\rm com} $.

\medskip

To simplify our analysis, we give the channel to whoever controlling
it. Therefore, it suffices to consider a two-party state in $H_A \otimes
H_B$.  Now, the fact that Bob has absolutely no information about the value
of $b$ implies that ideally the density matrix in his hand is independent of
the value of $b$.  That is to say that ${\rm Tr}_{A} \left(|0 \rangle_{\rm
com} \langle 0|_{\rm com} \right) = {\rm Tr}_{A} \left( |1 \rangle_{\rm com}
\langle 1 |_{\rm com} \right) $.  But then $|0 \rangle_{\rm com} $ and $|1
\rangle_{\rm com} $ of $H$ must have the same Schmidt decomposition (See for
example, the Appendix of Ref.~\cite{Hughston}.), namely:
\begin{equation}
|0 \rangle_{\rm com} = \left( \sum_k \sqrt{\lambda}_k 
| e_k \rangle_A \otimes | \phi_k \rangle_B \right) ,
\eqlabel{polar0}
\end{equation}
and
\begin{equation}
|1 \rangle_{\rm com} = \left( \sum_k \sqrt{\lambda}_k 
| e_k \rangle_A' \otimes | \phi_k \rangle_B \right)  ,
\eqlabel{polar1}
\end{equation}
 where $| e_k \rangle_A$ and $| e_k \rangle_A'$ are two orthonormal bases of
$H_A$ and $ | \phi_k \rangle_B$ is an orthonormal basis of $H_B$.  Notice
that $\lambda_k$ and $| \phi_k \rangle_B$ are the same in the above two
equations and that the only difference lies in Alice system ($| e_k
\rangle_A$ as opposed to $| e_k \rangle_A'$).  The key observation is that
these two states are related by a unitary transformation acting on $H_A$
alone!  Consequently, Alice can make a fake commitment and change the value
of $b$ in the beginning of step (c).

\medskip

For example, she may proceed as follows: First, Alice always takes $b=0$ in
step (a) and goes through step (b), the commit phase.  It is only in the
beginning of step (c), the opening phase,
that Alice decides on the actual value of $b$ that she wishes to open. If she
decides $b=0$ now, she can go through step (c) honestly. If she wishes to
change the value of $b$ from $0$ to $1$, she simply applies a unitary
transformation ($| e_k \rangle_A \to | e_k \rangle_A'$) to rotate the state
from $|0 \rangle_{\rm com}$ to $ |1 \rangle_{\rm com}$ before going through
step (c).  Since the unitary transformation acts on $H_A$ alone, Alice can
apply it without Bob's help.  Moreover, Bob clearly has no way of knowing
Alice's cheating.\footnote{What is the problem with quantum bit commitment?
Here is an analogy.  Suppose that there are two novels whose first halves are
the same, but the second halves are different. I give you only the first half
of one of the two novels and I tell you that I have committed to a particular
novel and that I cannot change it anymore. Will you trust me?  Of course
not. Since the first halves of the two novels are the same, no real
commitment has been made.  I am free to give you the second half of either
novel and claim that I have committed to either one all along. There is no
way for you to tell whether I am lying.}  In conclusion, provided that Alice
possesses quantum computers and quantum storage devices, our results show
that all quantum bit commitment schemes are insecure because Alice can cheat
successfully by using an EPR-type of attack.

\medskip

As remarked in the last subsection, this no-go theorem applies
to all bit commitment schemes---purely quantum, classical or quantum
but with classical measurements.
This is because any procedure of quantum bit commitment scheme
can be regarded as a scheme in which Bob does have
a quantum computer but fails to make full use of it.
By showing that Alice can cheat successfully {\it even if\/}
Bob makes full use of his quantum computer, we
break all bit commitment scheme---purely quantum, classical or quantum
but with classical measurements.

\subsection{Non-ideal bit commitment}

In our above discussion, we have assumed that the bit commitment scheme is
ideal in the sense that Bob has absolutely no information about the value of
$b$ at the end of step (b). This is the physical reason behind the
mathematical statement that $\rho_0^{\rm com}= {\rm Tr}_{A} \left( |0
\rangle_{\rm com} \langle 0|_{\rm com} \right) = {\rm Tr}_{A} \left( |1
\rangle_{\rm com} \langle 1 |_{\rm com}\right) = \rho_1^{\rm com}$.  (i.e.,
the two density matrices corresponding to the two cases $b=0$ and $b=1$ are
identical.)  However, in realistic applications, one might allow Bob to have
a very tiny amount of information about $b$ at that time.  It is intuitively
plausible that this is not going to change our conclusion. On the one hand,
if Bob has a large probability of distinguishing between the two states
corresponding to $b=0$ and $b=1$ at the end of step (b), the scheme is
inherently unsafe against Bob.  On the other hand, if Bob has a small
probability of distinguishing between the two states, then clearly, the
density matrices $\rho_0^{\rm com}= {\rm Tr}_{A} \left( |0 \rangle_{\rm com}
\langle 0|_{\rm com} \right)$ and $\rho_1^{\rm com}= {\rm Tr}_{A} \left( |1
\rangle_{\rm com} \langle 1 |_{\rm com} \right)$ must be close to each other
in some sense.  We have seen in the last subsection that when the two density
matrices are identical, Alice can always cheat successfully.  It is,
therefore, at least highly suggestive that, when the two density matrices are
only slightly different, Alice will have a probability close to $1$ of
cheating successfully.  The analysis in the next subsection shows that this
is indeed the case. Therefore, even non-ideal bit commitment schemes are
necessarily highly insecure.
 
\subsection{Fidelity}

In this subsection, following Mayers \cite{Mayers2}, we sketch the
mathematical proof of the insecurity of non-ideal quantum bit commitment
scheme.  Readers who are uninterested in mathematical details may skip this
subsection on first reading.  The price that they have to pay is to take
Eqs.~\eq{fidelity1} and \eq{fidelity3} for granted.

\medskip

First of all, the closeness between two density matrices $\rho_0$ and $\rho_1$
of a system $B$ can be described by {\em fidelity} \cite{Fuchs,Fuchs1,Jozsa},
which is defined as
 \begin{equation}
F(\rho_0, \rho_1) = {\rm Tr} \sqrt {\rho_1^{1/2} \rho_0 \rho_1^{1/2}} .
\eqlabel{fidelity}
\end{equation}
$0 \leq F \leq 1$. $F=1$ if and only if $\rho_0 = \rho_1$.
Returning to the case of non-ideal bit commitment that we have
been considering,
the fact that Bob has a small
probability for distinguishing between two states $\rho_0^{\rm com}$ and
$\rho_1^{\rm com}$
implies that the fidelity $F(\rho_0^{\rm com}, \rho_1^{\rm com})$ is
close to $1$. i.e.,
\begin{equation}
F(\rho_0^{\rm com}, \rho_1^{\rm com})= 1- \delta,
\eqlabel{largef}
\end{equation}
where $\delta$ is small.

\medskip

An alternative and equivalent definition of fidelity involves the concept of
{\em purification}.  Imagine another system $E$ attached to our given system
$B$.  There are many pure states $| \psi_0 \rangle$ and $| \psi_1 \rangle$ on
the composite system such that
 \begin{equation}
{\rm Tr}_E \left( | \psi_0 \rangle \langle \psi_0 | \right) = \rho_0 
\mbox{~~~~~~and~~~~~}
{\rm Tr}_E \left( | \psi_1 \rangle \langle \psi_1 | \right) = \rho_1.
\eqlabel{pure}
\end{equation}
 The pure states $| \psi_0 \rangle$ and $| \psi_1 \rangle$ are called the
purifications of the density matrices $\rho_0 $ and $\rho_1 $.  The fidelity
can be defined as
 \begin{equation}
F(\rho_0, \rho_1) = {\rm max} | \langle \psi_0|\psi_1  \rangle  |
\eqlabel{fidelity1}
\end{equation}
 where the maximization is over all possible purifications.

\medskip

Here are two useful remarks. Firstly, for any fixed purification of $\rho_1$,
there exists a maximally parallel purification of $\rho_0$ satisfying
Eq.~\eq{fidelity1}.\footnote{We thank R. Jozsa for a helpful discussion.}
Secondly, given any two purifications of $\rho_0$, there exists a local
unitary transformation which rotates one into the other.

\medskip

Let us go back to a non-ideal quantum bit commitment scheme.  We take $E$ to
be Alice's machine $A$.  Using the first remark, we find from
Eqs.~\eq{largef} and \eq{fidelity1} that, for the state $|1 \rangle_{\rm
com}$ which is a purification of $\rho_1^{\rm com}$, there exist a
purification $| \psi_0 \rangle_{AB}$ of $\rho_0^{\rm com}$ such that
 \begin{equation}
| \langle \psi_0| 1  \rangle_{com}  | =
F(\rho_0^{\rm com}, \rho_1^{\rm com})= 1- \delta .
\eqlabel{fidelity2}
\end{equation}

\medskip

The strategy of a cheating Alice is the same as in the ideal case.  She
always prepares the state $| 0 \rangle$ corresponding to $b=0$ in step (a)
and goes through step (b).  She decides on the value of $b$ she likes only in
the beginning of the step (c). If she chooses $b=0$, of course, she can just
follow the rule.  If she chooses $b=1$, she applies a unitary transformation
to obtain the state $| \psi_0 \rangle_{AB}$ in $H=H_A \otimes H_B $. The
existence of such a unitary transformation is guaranteed by the second
remark. After this unitary transformation,
she can continue to execute step~(c) of the protocol for $b =1$,
pretending that the total state is $|1 \rangle_{\rm com} $.

\medskip

If she had been honest, the state would have been $|1
\rangle_{\rm com} $ instead.  Since $| \psi_0 \rangle_{AB}$ and $|1
\rangle_{\rm com} $ are so similar to each other (See Eq.~\eq{fidelity2}.),
Bob clearly has a hard time in detecting the dishonesty of Alice.  Therefore,
Alice can cheat successfully with a very large probability.

\medskip

Yet another equivalent definition of the fidelity, which will be useful in
the next section, can be given in terms of positive-operator-valued-measures
(POVMs).  A POVM is a set $\{\hat{E_b}\}$ of positive operators (i.e.,
Hermitian operators with non-negative eigenvalues) that satisfy a sort of
completeness relation (i.e., $\sum_b \hat{E_b}$ equals the identity
operator).  A POVM simply represents the most general measurement that can be
performed on a system. More concretely, it is implemented by a) placing the
system in contact with an auxiliary system or ancilla prepared in a standard
state, b) evolving the two by a unitary operator, and c) performing an
ordinary von Neumann measurement on the ancilla.  In terms of POVMs, the
fidelity is defined as
 \begin{equation}
F(\rho_0, \rho_1) = {\rm min} \sum_b \sqrt{{\rm Tr} \rho_0 \hat{E_b}}
\sqrt{{\rm Tr}\rho_1 \hat{E_b}} ,
\eqlabel{fidelity3}
\end{equation}
 where the minimization is over all POVMs, $\{\hat{E_b}\}$.
Eq.~\eq{fidelity3} will be useful in the next section.

\section{Quantum coin tossing}

Suppose that Alice and Bob are having a divorce and that they are living far
away from each other. They would like to decide by a coin flip over the
telephone who is going to keep the house. Of course, if one of them is
tossing a real coin, there is no way for the other to tell if he/she is
cheating. Therefore, there must be something else that is simulating the coin
flip.  Just like bit commitment, coin tossing can be done in classical
cryptography either through trusted intermediaries or by accepting some
unproven computational assumptions.  The question is: Can quantum mechanics
help to remove those requirements? In other words, do coin tossing schemes
whose security is based solely on the law of quantum physics exist?

\medskip

Notice that a secure bit commitment protocol can be used trivially to
implement a secure coin tossing protocol\footnote{Alice chooses a bit and
commits it to Bob. Bob simply tells Alice his guess for her bit. Alice then
opens her commitment to see if Bob has guessed correctly.} but the converse
is not true. Coin tossing is a weaker protocol for which we have a weaker
result---{\em ideal\/} quantum coin tossing schemes do not exist.  We define
an ideal coin tossing scheme as one that satisfies the following
requirements:\footnote{We gratefully thank Goldenberg and Mayers for many
discussions which are very helpful for sharpening and clarifying our ideas.}

\medskip

1) At the end of the coin tossing scheme, there are three possible outcomes:
`$0$', `$1$' or `invalid'.

2) Both users know which outcome occurs.

3) If the outcomes `$0$' or `$1$' occur, Alice and Bob can be sure that they
occur with precisely the (non-zero) probabilities, say $1/2$ each, prescribed
by the protocol.

4) If both users are honest, the outcome `invalid' will never occur.

\medskip

In other words, in an {\em ideal\/} coin tossing scheme, both parties will
always agree with each other on the outcome. There is no room for
dispute. Also, cheating in an ideal coin tossing will only lead to a
non-vanishing probability for the occurrence of `invalid' as an outcome, but
will not change the relative probability of occurrence of `0' and `1'.  Most
coin tossing schemes are non-ideal.  However, any non-ideal quantum coin
tossing scheme can be regarded as an approximation to an ideal
scheme. Investigations of the ideal scheme may, therefore, shed some lights
on those more realistic, but non-ideal, ones.

\medskip

Of course, if one allows Alice and Bob to share {\it initially} some {\it
known} entangled state, for example, a singlet, they can trivially achieve
quantum coin tossing.  The problem is, however, that there is no way for
Alice and Bob to verify that, indeed, they share such a state.  If such a
state is prepared by a trusted third party, the scheme is no longer a
two-party protocol. If such a state is prepared by one of the two parties, he
or she may cheat. Therefore, one should always assume that initially Alice
and Bob share no entanglement.

\medskip

To show that ideal quantum coin tossing is impossible, we first prove the
following Lemma.

\medskip

{\it Lemma}: Given that Alice and Bob initially share no entangled quantum
states, they cannot achieve ideal quantum coin tossing without any further
communication between each other.

\medskip

{\it Proof}: An ideal coin tossing scheme will give Alice and Bob non-zero
mutual information.  However, without prior classical communication, the
maximal amount of mutual information that can be gained by Alice and Bob
through local operations on shared entangled quantum states is bounded by the
entropy of formation. In the absence of entanglement, they cannot share any
mutual information.  Hence, coin tossing without prior communication nor
shared entangled quantum states must be impossible.

\medskip

Now we come to the main theorem.

\medskip

{\it Theorem}: Given that Alice and Bob initially share no entangled quantum
states, {\em ideal} quantum coin tossing is impossible.

\medskip

{\it Proof}: The idea of our proof is simple. We prove by contradiction using
backward induction. Let us assume that an ideal quantum coin tossing can be
done with a fixed and finite number, $N$, rounds of communication between
Alice and Bob. We will prove that it can be done in $N-1$ rounds. By repeated
induction, it can be done without any communication between Alice and Bob at
all.  This is impossible because of the above Lemma.

\medskip

The induction step from $N$ rounds to $N-1$ rounds: Suppose that there exists
an ideal quantum coin tossing protocol which involves $N$ alternate rounds of
communication between Alice and Bob. We need to prove that an ideal quantum
coin tossing protocol with only $N-1$ rounds exists.  Let us concentrate on
the $N$-th round of the communication. Without much loss of generality,
assume that it is Alice's turn to send quantum signals through the channel
$C$ in the $N$-th round.  As this is the last round, by the definition of
ideal coin tossing, Alice can perform a measurement and determine the
outcome, $0$, $1$ or `invalid', before sending out the last round signals to
Bob. Notice that Alice should have no objection against eliminating the
$N$-th round altogether because she has nothing to gain in sending the last
round signals (other than convincing Bob of the outcome).  On the other hand,
Bob is supposed to learn the outcome of the coin tossing through the combined
state in $H_B \otimes H_C$.  However, Alice, who has already known the
outcome herself, may attempt to alter Bob's outcome by changing the mixed
state in $H_C$ that she is sending through the channel. This is essentially
the same strategy of cheating as in the case of quantum bit commitment
discussed earlier.

\medskip

For the three possible outcomes in Alice's measurement, $0$, $1$ and
`invalid', let us denote the corresponding density matrices in Bob's control
{\em before} the receipt of the $N$-th round signals by $\rho_0^B$,
$\rho_1^B$ and $\rho_{\rm invalid}^B$ respectively.  Alice's ability of
cheating successfully against an honest Bob depends on the values of the
fidelities $F(\rho_0^B, \rho_1^B)$, $F(\rho_0^B, \rho_{\rm invalid}^B)$ and
$F(\rho_1^B, \rho_{\rm invalid}^B)$.  Here for simplicity, we assume that
there is a {\it single} pure state corresponding to the outcome
`invalid'. However, our arguments are general.  For {\em ideal\/} quantum
coin tossing, we demand the probability of Alice cheating successfully should
be exactly zero. This implies, with the definition of fidelity in
Eq.~\eq{fidelity1}, that $F(\rho_0^B, \rho_1^B)=0$, $F(\rho_0^B, \rho_{\rm
invalid}^B)=0 $ and $F(\rho_1^B, \rho_{\rm invalid}^B)=0 $.  It then follows
from Eq.~\eq{fidelity3} that $\rho_0^B$, $\rho_1^B$ and $ \rho_{\rm
invalid}^B$ have orthogonal supports and can be completely distinguished from
one another even without the last round of transmission from Alice. Hence,
even Bob has nothing to gain from the last round of communication.  A
truncated ideal coin tossing scheme with only $N-1$ rounds of communication
must, therefore, be as secure as the original $N$-round scheme.  This
completes our inductive argument and we conclude that ideal quantum coin
tossing is impossible.  As noted in subsection 2.2, this ``no-go'' theorem
applies to {\it all} ideal coin tossing scheme including purely quantum,
purely classical and quantum but with classical measurements.

\medskip

Unlike quantum bit commitment, for quantum coin tossing, there is no simple
way to generalize our proof of the impossibility of the ideal scheme to the
non-ideal ones. This is surprising because no such distinction has been
previously noted in the literature. As far as we know, all previously
proposed quantum coin tossing schemes are based on quantum bit commitment
schemes.  The security of non-ideal quantum coin tossing should be an
important subject for future investigations.  We hope that our investigation
for the ideal case will shed light on the subtleties in the non-ideal case.

\section{A constraint on two-party secure computation}

Let us consider the issue of two-party secure computation in a more general
setting.  The idea of two-party secure computation is the following: Alice
has a secret $x$ and Bob has another secret $y$.  Both would like to know the
result $f(x,y)$ at the end of a computation and be sure that the result is
correct.  However, neither side wishes the other side to learn more about its
own secret than what can be deduced from the output $f(x,y)$. As mentioned
earlier, classical cryptographic schemes can implement two-party secure
computation at the cost of introducing trusted intermediaries or accepting
unproven cryptographic assumptions.  Our results in the last two sections
strongly suggest that, in principle at least, quantum cryptography would not
be useful for getting rid of those requirements in two-party secure
computation.  Even if quantum mechanics does not help, one may ask if there
is {\em any} way of implementing a two-party computation that is secure from
an information-theoretic point of view? In particular, if quantum mechanics
turned out to be wrong and were replaced by a new physical theory, would it
be conceivable that two-party secure computation can be done in this new
theory?  Here we argue that if Alice and Bob are shameless enough to declare
their dishonesty and stop the computation whenever one of them has a
slightest advantage over the other in the amount of mutual information he/she
has on the function $f(x,y)$, a two-party secure computation can never be
implemented.

\medskip

For simplicity, let us normalize everything and assume that initially both
Alice and Bob have no information about $f(x,y)$ and at the end of the
computation, both have $1$ bit of information about $f(x,y)$. Let us suppose
further than Alice and Bob are unkind enough to stop the computation whenever
one of them has an $\epsilon$ bit of information more than the other.  Any
realistic scheme must involve a finite number say $N$ alternate rounds of
communication between Alice and Bob.  An analogy is that two persons, Alice
and Bob, are walking in discrete alternate steps from the starting point $0$
to the finishing line set at $1$.  Altogether $N$ steps are made and it is
demanded that Alice and Bob will never be separated from each other for more
than a distance $\epsilon$.\footnote{Actually, there is a minor subtlety in
quantum cryptography.  Each time when one user, say Alice, advances, the
other user, say Bob, may slip backwards. The point is: the quantum
``no-cloning theorem'' states that quantum signals cannot be copied. When Bob
sends signals to Alice, he loses control over the signals that he sends.  In
other words, Bob's available information tends to decrease whenever he sends
signals to Alice.}  Clearly, this implies $N \epsilon \geq 1$ or $N \geq
1/\epsilon$.  Therefore, the smaller the tolerable relative informational
advantage $\epsilon$ is, the larger the number of rounds of communication $N$
is needed.  Notice that the constraint $N \epsilon \geq 1$ applies to {\em
any} two-party secure computation scheme. In particular, it remains valid
even if quantum mechanics is wrong.

\medskip

It may also be of some interest to speculate that a similar inequality
$N \epsilon \geq 0(1) $ may hold for non-ideal quantum coin tossing schemes
where $N$ is the number of rounds of communication and $\epsilon$ is
the probability that a user cheats successfully. Consequently, as
$\epsilon \to 0$, $N \to \infty$ and ideal quantum coin tossing
with finite rounds of communication becomes impossible.

\section{Summary}

We have proven a no-go theorem for unconditionally secure quantum bit
commitment. We emphasize its generality.
This result applies to all bit commitment schemes---fully quantum,
classical or quantum but with measurements. The basic problem
is that the users can cheat using entanglement to perform an EPR-type of
attack. This attack is a re-incarnation of the famous Einstein-Podolsky-Rosen
paradox. While this basic attack was well-known, its generality was
not widely appreciated until recently \cite{Mayers2,Lo2}.

Our simple proof of the no-go theorem
uses a unitary description of the execution of the protocol.
The point is the following:
quantum mechanics must provide the simplest and
most fundamental description of any protocol based on quantum physics.
In particular, there is no need to give measurements and classical
communications in the intermediate steps any special treatment.
Indeed, we argue that by including all the ancillas' states into
consideration and considering the quantum
mechanics for the enlarged system, one can describe any quantum cryptographic protocol
in a simple unitary setting.
Put it another way: Any procedure followed by Bob is equivalent to
the situation when Bob has a quantum computer but fails to make full use of it.
Therefore, by showing that Alice can cheat almost
always successfully against Bob {\it even if\/} Bob makes
full use of his quantum computer, it is clear that this `sure-win' strategy
will defeat any strategy followed by Bob. Consequently, quantum
bit commitment is impossible.

Our results totally contradict well-known claims of
unconditionally secure schemes in the literature, whose analyses on EPR attack were
flawed, and provide strong evidence against the security of quantum
cryptography in ``post-cold-war'' applications, at least {\em in
principle}. The early optimism in the subject is, therefore,
misplaced. Nevertheless, quantum bit commitment schemes that are secure {\em
in practice} may still exist\footnote{We thank C. H. Bennett for a discussion
about this point.} because it is notoriously difficult for cheaters with
current technology to store quantum signals for an arbitrary length of
time. (This can, in principle,
be overcome by quantum error correction \cite{Shor1,Steane}
which lengthens the decoherence time to arbitrarily long.)
A more serious consideration is the following. 
In order to cheat
successfully, a cheater generally needs a quantum computer, but such a computer is
not yet available with current technology.
(Errors in quantum computing can, in principle, be surmounted if a cheater
performs quantum computation in a {\it fault tolerant} \cite{Shor2,Preskill}
manner.)
Therefore, we can trade the
classical computational complexity assumptions with quantum
computational complexity assumptions. This subject deserves further investigations.

Notice also that while quantum computers can crack conventional cryptography
retrospectively, they cannot do so against quantum cryptography.
In this aspect, before a fully operating quantum computer is built,
quantum cryptography does have an advantage over conventional cryptography.

An important unsolved problem is the powers and limitations of quantum
cryptography. For example, is non-ideal quantum coin
tossing possible? How secure
is quantum money? Finally, we remark that, thanks to the quantum ``no-cloning''
theorem, the security of quantum {\em key distribution}
\cite{Mor,Deutsch,Mayers1,Mayers3} is widely accepted and quantum
cryptography is useful at least for this purpose. We expect quantum key
distribution to remain a fertile subject for years to come.

\medskip

{\it Notes added}: The impossibility of quantum bit commitment has also been
demonstrated by Mayers \cite{Mayers4}. One of us (H.-K. L) has recently
proven two new no-go theorems for other quantum cryptographic schemes, namely
quantum ``one-out-of-two
oblivious transfer'' and quantum ``one-sided'' two-party secure computations
\cite{Lo3}. There is also a survey paper on the insecurity of quantum
cryptographic schemes\cite{Lo4} by us.

\section{Acknowledgments}

Numerous helpful discussions with M. Ardehali, C. A. Fuchs, J. Kilian,
J. Preskill, M. H. Rubin, P. W.
Shor, L. Vaidman, A. Wigderson, F. Wilczek, and A. Yao
are acknowledged. We are very grateful to L. Goldenberg and D. Mayers for
pointing out a crucial error in the discussion of quantum coin tossing in an
earlier version of this manuscript and for many discussions which are very
helpful for sharpening and clarifying our ideas.  We acknowledge receipt of
preprints on the impossibility of quantum bit commitment from D. Mayers and
on coin tossing from L. Goldenberg.  We also thank C. H. Bennett,
D. P. DiVincenzo, R. Jozsa and T. Toffoli for generous advice and useful
comments. This research was supported by DOE grant DE-FG02-90ER40542

{\it Notes added}:
More recent discussions with numerous other
colleagues including A. Bodor, G. Brassard, C. Cr\'{e}peau, A. K. Ekert,
D. Gottesman, W. Y. Hwang,
A. Kent, S. Popescu and T. Spiller are also gratefully acknowledged.

\end{document}